\documentstyle[twocolumn,eqsecnum,aps,epsf,prb,amsmath,amssymb]{revtex}

\newcommand{\rd}{d}
\newcommand{\ri}{i}
\newcommand{\mtau}{\mbox{\boldmath$\tau$}}

\begin{document}


\title{Charge fluctuation between even and odd states of a superconducting island
}
\author{Yasuhiro Utsumi,
Hiroshi Imamura, 
Masahiko Hayashi, 
and 
Hiromichi Ebisawa}
\address{
Graduate School 
of Information Sciences,
Tohoku University, Sendai 980-8579, Japan
}
\date{\today}
\maketitle
\begin{abstract}
We theoretically investigate effects of quantum fluctuation between
an even and an odd charge state on the transport properties of a
normal-superconducting-normal single-electron tunneling transistor. 
The charge fluctuation is discussed beyond the orthodox theory. 
We find that the charging energy renormalization enhances the parity
effect: The Coulomb blockade regime for the odd state is reduced and that
for the even state is widened.   We show that the renormalization
factor can be obtained experimentally and that the renormalization effect 
is weakened by applied bias voltage.
\end{abstract}
\vskip1pc

%

\section{Introduction}

Quantum fluctuations play important roles on transport properties of
mesoscopic systems at low temperatures.
Recently the charge fluctuation in Coulomb islands 
has attracted much attention. 
There have been much development in the strong tunneling regime. 
A theoretical prediction, a renormalization of the conductance
and charging energy $E_C$\cite{Falci,Schoeller1,Konig,Schoeller2,Golubev}, 
has been confirmed experimentally\cite{Joyez,Chouvaev}. 

When the island is made of superconductor, the \lq\lq parity effect"
plays an important role on transport properties
\cite{RBT,Fazio,Utsumi1,Hekking1,Hekking2}. 
When $E_C$ exceeds the superconducting gap $\Delta$, the parity effect 
appears in the period of Coulomb oscillation where 
intervals are either elongated or shortened for even 
or odd occupancy of electrons respectively. 
At a resonance, an even and an odd state are degenerate and
an unpaired \lq\lq odd" electron
makes dominant contribution to the tunneling current. 
For $E_C<\Delta$, odd states are no longer stable for every gate voltage and 
thus 2$e$-periodic Coulomb oscillation appears\cite{Hekking1}. 

Therefore, it is an intriguing question to ask how the
parity effect with charge fluctuation affect on transport properties. 
In this paper,
we investigate the conduction properties of a superconducting island
for $E_C>\Delta$ with a special attention to the charge fluctuation
between an even and an odd state. 
For simplicity we limit ourselves to the discussion at zero temperature limit
and neglect the interference effect around tunnel barriers\cite{Hekking2}. 
We find that the interval of conductance peaks related to 
the Coulomb blockade (CB) regime for the odd state is shortened 
and that related to the even state is widened 
as a result of the charging energy renormalization. 
We show that the renormalization factor can be obtained experimentally and 
that the renormalization effect is weakened by applied bias voltage. 

The outline of this paper is as follows. 
In Sec. \ref{sec:model}, we introduce the model 
and derive the generating functional in the path-integral representation. 
We also propose the approximate generating functional Eq. (\ref{eqn:WRTA}). 
In Sec. \ref{sec:average}, we derive the current expression by using the
functional derivative. 
In Sec. \ref{sec:results} we show numerically evaluated results and 
have some discussions. 
Section \ref{sec:summary} summarizes our results.

\section{model and generating functional}
\label{sec:model}

\begin{figure}[ht]
\epsfxsize=\linewidth
\epsffile{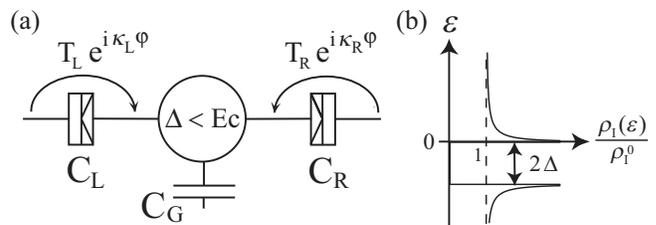}
\caption{
(a) The equivalent circuit of the NSN transistor. 
(b) The normalized density of states in the superconducting island. 
}
\label{fig:system}
\end{figure}

Figure \ref{fig:system}(a) shows an equivalent circuit of 
a normal-superconducting-normal (NSN) transistor. 
A superconducting island exchanges quasiparticles (QPs) with a left 
(right) lead via a small tunnel junction characterized by 
the tunnel matrix element $T_{\rm L(R)}$ and a capacitor $C_{\rm L(R)}$ 
and is coupled to a gate via a capacitor $C_{\rm G}$. 
In the following discussion, we limit ourselves to the symmetric case, $C_{\rm L}=C_{\rm R}$ and $T_{\rm L}=T_{\rm R}$. 
We use the two-state model to describe the strong Coulomb interaction. 
We assume that there are even (odd) number of electrons in a 
charge state $\left|0 (1)\right>$. 

The total Hamiltonian $H$ is given by the sum of $H_0$ and 
a tunneling Hamiltonian $H_{\rm T}$, 
which is adiabatically switched on at remote past and off at distant future. 
$H_0$ is defined as
\begin{equation}
H_0
=
\sum_{{\rm r}, k, n}
\varepsilon_{{\rm r} k}
\:
a_{{\rm r} k n}^{\dag} 
a_{{\rm r} k n}
+
\Delta_0 \frac{\sigma_{z}+1}{2}
+
E_C (Q_{\rm G}/e)^2, 
\label{eqn:H0}
\end{equation}
where $a_{{\rm r} k n}$ is the annihilation operator of a QP in the lead 
(${\rm r=L,R}$) or in the island (${\rm r=I}$) with wave vector $k$ and 
transverse channel $n$ which includes spins. 
$\sigma$ is the effective spin-1/2 operator. 
The second term of Eq. (\ref{eqn:H0}) describes the energy difference 
between two charge states. 
The excitation energy $\Delta_0=E_C(1-2 Q_{\rm G}/e)+\Delta$ is related to a gate charge $Q_{\rm G}$. 
The third term of Eq. (\ref{eqn:H0}) represents the charging energy 
for the even state. 

The density of states (DOS) for normal metal leads is approximated as 
constant: 
$\rho_{\rm r}(\varepsilon)
=\sum_k \delta(\varepsilon-\varepsilon_{{\rm r} \, k})
=\rho_{\rm r}$ ($\rm r=L,R$). 
Figure \ref{fig:system}(b) shows the normalized DOS in 
the island $\rho_{\rm I}(\varepsilon)/\rho^0_{\rm I}$ where $\rho^0_{\rm I}$ 
is DOS for the normal state. 
We assume that an unpaired electron occupies 
the lowest energy level above the superconducting gap in the limit of 0 K: 
\begin{eqnarray}
\rho_{\rm I}(\varepsilon)
&=&
\frac{\delta(\varepsilon-0)}{2}
+
\left\{
\begin{array}{ll}
\frac{\rho^0_{\rm I}|\varepsilon+\Delta|}
{\sqrt{(\varepsilon+\Delta)^2-\Delta^2}} & (|\varepsilon+\Delta|>\Delta) \\
0 & (|\varepsilon+\Delta| \le \Delta),
\end{array}
\right.
\label{eqn:s-DOS}
\end{eqnarray}
where we consider the condition that the charge and the spin 
relaxation in the island are enough fast and 
occupation probabilities for the up- and the down-spin unpaired 
electron are both $1/2$. 
We expect that the most dominant part of the superconducting 
correlation effects can be included via the shape of DOS, 
because any other process than the QP tunneling, 
such as the Andreev reflection process, requires higher energy 
$\sim E_C$. 

The tunneling Hamiltonian 
\begin{equation}
H_{\rm T}
=
\sum_{{\rm r=L,R}}
\sum_{k, k', n}
T_{\rm r} e^{\ri \varphi_{\rm r}}
\:
a_{{\rm I} k n}^{\dag} 
a_{{\rm r} k' n}
\sigma_{+}
+
{\rm h. c.},
\end{equation}
describes the electron tunneling across the junctions and simultaneous flip of the effective spin. 
The phase difference between the lead ${\rm r}$ and the 
island is written as 
$\varphi_{\rm r}=\kappa_{\rm r} \varphi$. 
$\varphi$ is the phase difference between two leads and is regarded 
as a source field. 
$\kappa_{\rm L(R)}=1/2(-1/2)$ characterizes the voltage drop between 
the left (right) lead and the island.

The Hamiltonian includes the effective spin-1/2 operator $\sigma$ and 
thus the Wick's theorem for fermions or bosons
cannot be used. 
To circumvent this drawback, we employ the drone-fermion representation
\cite{Isawa,Spencer}, 
which is the map of $\sigma$ 
onto two fermion operators $c$ and $d$ as 
$\sigma_{+}=c^{\dag} \phi$ and $\sigma_{z}=2c^{\dag}c-1$, 
where $\phi=d^{\dag}+d$ is a Majorana fermion operator. 
In our calculation, we reformulate Ref. \cite{Isawa} based on 
the Schwinger-Keldysh approach\cite{Keldysh,Chou} 
which enables us to obtain the average current, the current noise 
and any higher order moments systematically by 
the functional derivative in terms of the phase difference, 
{\it i.e.} a vector potential\cite{Kamenev,Gutman1}.

\begin{figure}[ht]
\epsfxsize=.7 \linewidth 
\centerline{\epsffile{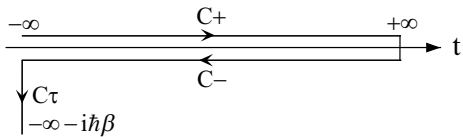}}
\caption{
The closed time-path going from $-\infty$ to $\infty$ ($C_{+}$), 
going back to $-\infty$ ($C_{-}$), connecting 
the imaginary time path $C_{\tau}$ and closing at $t=-\infty-\ri \hbar \beta$. 
}
\label{fig:ClosedTimePath}
\end{figure}

The action can be obtained from the total Hamiltonian $H$ 
by following the standard procedure\cite{Babichenko}:
\begin{eqnarray}
&S&[c^*,c,d^*,d,
a_{{\rm r} k n}^*,a_{{\rm r} k n}]
\nonumber \\
&=&
\int_C \rd t 
\{
c(t)^* (\ri \hbar \, \partial_t-\Delta_0) c(t)
+
\ri \hbar \, d(t)^* \partial_t d(t)
\nonumber
\\
&+&
\sum_{{\rm r},k, n}
a_{{\rm r} k n}(t)^* (\ri \hbar \, \partial_t-\varepsilon_{{\rm r} k}) a_{{\rm r} k n}(t)
\nonumber
\\
&+&
\sum_{\stackrel{\scriptstyle{\rm r=L,R}}{k, k', n}}
T_{\rm r}
{\rm e}^{\ri \varphi_{\rm r}(t)}
a_{{\rm r} k n}(t)^*
a_{{\rm I} k' n}(t)
\sigma_{+}(t)
+{\rm h. c.}
\},
\end{eqnarray}
where the derivative and the integration are performed along the closed 
time-path $C$ consisting of the forward branch $C_{+}$, 
the backward branch $C_{-}$, and the imaginary time path 
$C_{\tau}$\cite{Kiselev}(Fig. \ref{fig:ClosedTimePath}). 
The degrees of freedom for $\varphi$ are duplicated, {\it i.e.}, we can define 
$\varphi_{+}$ and $\varphi_{-}$ on the forward and the backward branch, 
respectively.
$a_{{\rm r} k n}$, $c$ and $d$ are Grassmann variables which satisfy 
the anti-periodic boundary condition. 
By tracing out QP degrees of freedom\cite{Golubev}
(Appendix. \ref{appendix:functionalintegration}), 
we obtain the following effective action for the $c$- and $d$-field: 
\begin{eqnarray}
&S&[c^*,c,d^*,d]
\nonumber \\
&=&
\int _C \rd t 
\{
c(t)^* (\ri \hbar \, \partial_t-\Delta_0) c(t)
+
\ri \hbar \, d(t)^* \partial_t d(t)
\}
\nonumber \\
&+&
\int _C \rd t_1 \rd t_2
\{
\sigma_{+}(t_1)
\alpha(t_1,t_2)
\sigma_{-}(t_2)
+
O(T_{\rm r}^4)
\}, 
\label{eq:effective-action}
\end{eqnarray}
where trivial constants are omitted. 
The second integral of Eq. (\ref{eq:effective-action}) describes the 
tunneling process. 
$\alpha=\sum_{\rm r=L, R} \alpha_{\rm r}$ is a particle-hole Green function 
(GF), written as
\begin{equation}
\alpha_{\rm r}(t,t')
=
- \ri \hbar
N_{\rm ch}
T_{\rm r}^2
\sum_{k,k'}
g_{{\rm r} k}(t,t') \,
g_{{\rm I} k'}(t',t)
\,
{\rm e}^{\ri \kappa_{\rm r}(\varphi(t)-\varphi(t'))}.
\label{eqn:gph}
\end{equation}
Here $N_{\rm ch}$ is the number of transverse channels and 
$g_{{\rm r} k}$ is a free QP GF in the lead ${\rm r}$ $({\rm r=L,R})$
or in the island $({\rm r=I})$. 
The inverse of $g_{{\rm r} k}$ is defined as 
\begin{equation}
g_{{\rm r} k}^{-1}(t,t')
=
(\ri \hbar \, \partial_t - \varepsilon_{{\rm r} k})
\,
\delta(t,t'),
\label{eqn:gk}
\end{equation}
where $\delta$-function is defined on $C$ and 
$g_{{\rm r} k}$ satisfies the anti-periodic boundary condition:
$g_{{\rm r} k}(t,-\infty \in C_{+})=-g_{{\rm r} k}(t,-\ri \hbar \beta-\infty)$.

In the wide junction limit, $N_{\rm ch} \rightarrow \infty$, the terms higher than $T_{\rm r}^4$, which describes elastic co-tunneling processes, 
can be neglected\cite{Konig}. 
The  particle-hole GF $\alpha$ describes tunneling and relaxation process 
related to the lowest order sequential tunneling. 
However, by tracing out $c$- or $d$-field, a number of $\alpha$ are coupled. 
Therefore, the effective action describes the higher order inelastic 
co-tunneling processes, too. 

Next we trace out drone-fermion fields. 
Firstly, we trace out the degrees of freedom for $c$-field. 
The resulting effective action for $d$-field includes the 
many-body interaction, 
and cannot be solved exactly. 
Thus we introduce a liner source term $\int_C \rd t_1 J(t_1) \phi(t_1)$, 
where $J$ is a Grassmann variable defined on $C$. 
By tracing out $d$-field degrees of freedom, 
we obtain the generating functional as
\begin{eqnarray}
Z
&=&
\exp \left(-\sum_n \frac{(\ri \hbar)^{2 n}}{n}
{\rm Tr} \left[\left(
g_c \frac{\delta}{\delta J} \alpha \frac{\delta}{\delta J} \right)^n \right] 
\right)
\nonumber
\\
&\times&
\left.
\exp \left(-\frac{1}{2 \ri \hbar}
\int_C \rd t_1 \rd t_2 \, J(t_1) g_{\phi}(t_1,t_2) J(t_2) \right) \right|_{J=0}
\nonumber \\
&\times&
{\rm e}^{{\rm Tr} \ln \left[ g_c^{-1} \right]},
\label{eqn:startZ}
\end{eqnarray}
where trace is performed over $C$ and products in the 
trace represent the integration along $C$. 
We omit the factor $2$ which is the partition function of $d$-field. 
Here the $c$-field and the $d$-field GFs are defined as 
\begin{eqnarray}
g_c^{-1}(t,t')
&=&
(\ri \hbar \, \partial_t -\Delta_0) \, \delta(t,t'),
\label{eqn:GFc}
\\
g_{\phi}^{-1}(t,t')
&=&
\ri \hbar \, \partial_t \delta(t,t')/2. 
\label{eqn:GFd}
\end{eqnarray}
Both GFs satisfy the anti-periodic boundary condition.

The generating functional for connected GF, 
$W=-\ri \hbar \ln Z$, 
is evaluated by performing the perturbation series expansion in powers of 
$\alpha$, namely the dimensionless junction conductance
$\alpha_0=\sum_{\rm r=L,R} \alpha_{\rm r}^0$ 
where 
$\alpha_{\rm r}^0=N_{\rm ch} T_{\rm r}^2 \rho^0_{\rm I} \rho_{\rm r}$. 
For example, the first order contribution is written as
$W^{(1)}=\ri \hbar \, {\rm Tr} \left[ g_c \Sigma_c \right]$,
where the self-energy $\Sigma_c=\sum_{\rm r=L,R} \Sigma_{\rm r}$ is defined as
\begin{eqnarray}
\Sigma_{\rm r}(t,t') &=& -\ri \hbar \, g_{\phi}(t',t) \, \alpha_{\rm r}(t,t').
\label{eqn:selfc}
\end{eqnarray}
A finite order contribution causes a divergence of the physical quantity, 
such as the average charge, 
at the degeneracy point $\Delta_0=0$ (Appendix. \ref{appendix:charge1st}). 
To regularize the divergence, one must take infinite orders into account. 
The most simple way is to sum up order-$n$ $c$-field 
corrections $(g_c \Sigma_c)^n$. 
This strategy is also adopted by Isawa {\it et. al.} in Ref. \cite{Isawa}, 
though their formulation is different from ours. 
The resulting approximate generating functional is expressed as 
\begin{equation}
\bar{W} = - \ri \hbar {\rm Tr}[\ln G_c^{-1}],
\label{eqn:WRTA}
\end{equation}
where $G_c$ is the full $c$-field GF whose inverse is 
defined as
\begin{eqnarray}
G_c^{-1}(t,t') &=& g_c^{-1}(t,t')-\Sigma_c(t,t'). 
\label{eqn:fullgc}
\end{eqnarray}
As we show in the next section, the approximate generating functional
formally reproduces the result of the resonant tunneling approximation
(RTA)\cite{Schoeller1,Isawa}. 

\section{average current}
\label{sec:average}

The tunneling current is obtained by functional derivative of the generating functional $\bar{W}$ with respect to the phase difference\cite{Kamenev} as
$
I
=
\left.
(e/\hbar)
\delta \bar{W}
/
\delta \varphi_{\Delta}(t)
\right|_{
\varphi_{\Delta}=0
}
=\sum_{\rm r=L,R}
\kappa_{\rm r}
I_{\rm r}.
$
The relative coordinate $\varphi_{\Delta}=\varphi_{+}-\varphi_{-}$ is a fictitious variable and must be 0. 
The center-of-mass coordinate, $\varphi_c=(\varphi_{+}+\varphi_{-})/2$, is a 
physical variable fixed at $eV t /\hbar$. 
By regarding $\varphi_{\rm r}$ as the formally independent variable, 
the tunneling current through the junction ${\rm r}$ 
,$I_{\rm r}$, is expressed as
\begin{eqnarray}
&I_{\rm r}&(t)
=
\frac{e}{\hbar}
\left. 
\frac{\delta \bar{W}}{\delta \varphi_{{\rm r} \, \Delta}(t)} 
\right|_{\varphi_{\Delta}=0}
\nonumber
\\
&=&
-e
\left.
{\rm Tr}
\left[
G_c
\frac{\delta \varphi_{\rm r}}{\delta \varphi_{{\rm r} \, \Delta}(t)}
\Sigma_{\rm r}
-
G_c
\Sigma_{\rm r}
\frac{\delta \varphi_{\rm r}}{\delta \varphi_{{\rm r} \, \Delta}(t)}
\right]
\right|_{\varphi_{\Delta}=0}
\label{eqn:trace1}.
\end{eqnarray}
Here, we pay attention to the fact that only the self-energy 
depends on the phase difference. 
Next we project the time defined along $C$ to the real axis. 
As the tunneling Hamiltonian is turned on and off adiabatically,
the particle-hole GF is zero on the imaginary time path $C_{\tau}$. 
Thus, Eq. (\ref{eqn:trace1}) is rewritten as, 
\begin{eqnarray}
e
{\rm Tr}
\left.
\left[
\tilde{G}_c
\mtau^1
\tilde{\Sigma}_{\rm r}
\frac{\delta \tilde{\varphi}_{\rm r}}
{\delta \varphi_{{\rm r} \, \Delta}(t)}
-
\tilde{G}_c
\frac{\delta \tilde{\varphi}_{\rm r}}
{\delta \varphi_{{\rm r} \, \Delta}(t)}
\tilde{\Sigma}_{\rm r}
\mtau^1
\right]
\right|_{\varphi_{\Delta}=0},
\label{eqn:trace2}
\end{eqnarray}
where the trace is performed in the $2 \times 2$ Keldysh space and products 
represent the integration along the real axis. 
$\mtau^{s} \: (s=0,1,2,3)$ is the Pauli matrix in the Keldysh space. 
Here GF and the phase difference denoted with 
tilde are the $2\times2$ matrix for GF and that for the scalar variable in the 
{\it physical representation}\cite{Chou}: 
\begin{equation}
\tilde{G_c}
=
\left(
\begin{array}{cc}
0 & G_c^A \\
G_c^R & G_c^K
\end{array}
\right),
\,
\tilde{\Sigma_c}
=
\left(
\begin{array}{cc}
0 & \Sigma_c^A \\
\Sigma_c^R & \Sigma_c^K
\end{array}
\right),
\label{eqn:physical}
\end{equation}
\begin{equation}
\tilde{\varphi}_{\rm r}
=
Q
\left(
\begin{array}{cc}
\varphi_{\rm r \, +} & 0 \\
0 & -\varphi_{\rm r \, -}
\end{array}
\right)
Q^{\dagger}
=
\varphi_{\rm r \, c} \mtau^1+\varphi_{\rm r \, \Delta} \frac{\mtau^0}{2},
\label{eqn:physicalphi}
\end{equation}
where GFs with superscripts, $A$, $R$ and $K$, 
represent the retarded, the advanced and the Keldysh component, 
respectively. 
The practical calculations of these components 
are shown in Appendix. \ref{appendix:greenfunction}. 
The matrix $Q$ is the Keldysh rotator\cite{Chou}. 
From Eq. (\ref{eqn:physicalphi}), we derive the useful relation for the 
functional derivative technique: 
$
\delta \tilde{\varphi}_{\rm r}(t')/\delta \varphi_{{\rm r} \, \Delta}(t)
=\mtau^0\delta(t-t')/2
$.
%
By using the property of GF in the physical representation 
$\tilde{G}(t,t')^{\dagger}=-\mtau^3 \tilde{G}(t',t) \mtau^3$, 
and that of a Pauli matrix $\mtau^3 \mtau^1 \mtau^3=-\mtau^1$, 
we can see that the second term of Eq. (\ref{eqn:trace2}) is 
minus the complex conjugate of the first term. 
By performing the Fourier transformation, Eq. (\ref{eqn:trace2}) 
becomes as
\begin{eqnarray}
& &2e {\rm Re}
\left.
\int \rd t_1
{\rm Tr} \left[
\tilde{G}_c(t,t_1) \mtau^1 \tilde{\Sigma}_{\rm r}(t_1,t) \frac{\mtau^0}{2}
\right]
\right|_{\varphi_{\Delta}=0}
\nonumber \\
&=&
2e {\rm Re}
\int \frac{\rd \varepsilon}{h}
{\rm Tr} \left[
\tilde{G}_c(\varepsilon) \mtau^1
\tilde{\Sigma}_{\rm r}(\varepsilon) \frac{\mtau^0}{2} 
\right]
\nonumber \\
&=&
e
\int
\frac{\rd \varepsilon}{h}
\frac{
\Sigma_{\rm r}^K(\varepsilon) G_c^C(\varepsilon)
}{2}
-
(C \leftrightarrow K),
\label{eqn:Iphysical}
\end{eqnarray}
where $G_c^C=G_c^R-G_c^A$, {\it etc.} 
By using the expressions for the full $c$-field GF 
(Eqs. (\ref{eqn:cGC}) and (\ref{eqn:cGK})) 
and those for the self-energy (Eqs. (\ref{eqn:selfenergyR}) 
and (\ref{eqn:selfenergyK})), 
we obtain the expression for the average current in the limit of 0 K: 
\begin{eqnarray}
I
&=&
-
\frac{G_{\rm K}}{e}
\int^{eV/2}_{-eV/2} \rd \varepsilon
\frac{\alpha_{\rm L}^K(\varepsilon) \alpha_{\rm R}^K(\varepsilon)}
{\alpha^K(\varepsilon)}
2 \ri {\rm Im} G_c^{R}(\varepsilon),
\label{eq:current}
\end{eqnarray}
where $G_{\rm K}=e^2/h$ is the conductance quantum and 
$\alpha_{\rm r}^K$ is the Keldysh component of the particle-hole GF
(Eq. (\ref{eqn:alphaK})). 
The particle-hole GF of the superconducting island in Eq. (\ref{eq:current}) is given by 
$\alpha^K_{\rm r}(\varepsilon)
=-2 \pi \ri \, \alpha_{\rm r}^0 
\, |\rho(\delta \varepsilon^{\rm r}-\Delta)|$ where 
$\delta \varepsilon^{\rm r}=\varepsilon-\kappa_{\rm r} eV$.  
The spectral function $\rho(\varepsilon)$ is given by 
\begin{eqnarray}
\rho(\varepsilon)
&=&
\left\{
\begin{array}{ll}
-\sqrt{\varepsilon^2-\Delta^2} & (-D<\varepsilon \leq -\Delta) \\
1/2 \rho_{\rm I}^0 & (|\varepsilon| < \Delta) \\
1/2 \rho_{\rm I}^0 +\sqrt{\varepsilon^2-\Delta^2} & 
(\Delta \leq \varepsilon<D), 
\end{array}
\right.
\label{eqn:spectralfunction}
\end{eqnarray}
where the high energy cut-off $D$ $(=E_C \gg \Delta)$ is 
introduced\cite{Schoeller1}. 
The imaginary part of the retarded $c$-field GF 
$G_c^R(\varepsilon)=1/(\varepsilon-\Delta_0-\Sigma_c^R(\varepsilon))$ 
describes the excitation spectral density of the charge state. 
The self-energy of $c$-field is expressed by 
$
\Sigma_c^R(\varepsilon)
=
\sum_{\rm r=L,R} \alpha_{\rm r}^0 R(\delta \varepsilon^{\rm r}-\Delta)
-
\ri
\gamma(\varepsilon), 
$
where the function $R$ is written as 
\begin{eqnarray}
R(\varepsilon)
&\sim&
-2 \: \varepsilon \ln 
\left(2D/\Delta\right)
+
2 \sqrt{\varepsilon^2-\Delta^2}
\nonumber
\\
&\times&
{\rm Re} \left[
\tanh^{-1}
\left(
\frac{D \varepsilon}{\Delta^2-\sqrt{D^2-\Delta^2} 
\sqrt{\varepsilon^2-\Delta^2}}
\right)+h(\varepsilon)
\right]
\nonumber \\
&-&
\ln(|(\varepsilon-D)/(\varepsilon+\Delta)|)/\rho_{\rm I}^0,
\label{eq:self-energy}
\end{eqnarray}
where 
$h(\varepsilon)=0$ 
for $|\varepsilon|<\Delta$ and $h(\varepsilon)=\tanh^{-1}(D/\varepsilon)$ 
otherwise. 
The imaginary part of the self-energy 
$\gamma(\varepsilon)=- {\rm Im} \left[ \alpha^K(\varepsilon) \right] /2$ 
represents the life-time broadening of the charge state caused by 
dissipative charge fluctuation. 
The charge fluctuation described by $\alpha^K$ is suppressed in 
the low energy range, $0<\varepsilon<2 \Delta$ $\ll D$, 
because for the odd state, there are no states for QPs 
in the range $-2 \Delta<\varepsilon<0$ as shown in Fig. \ref{fig:system}(b). 
The asymmetry in particle-hole GF $\alpha^K(\varepsilon)$ causes 
the renormalization of the peak position of ${\rm Im}G_c^R(\varepsilon)$. 
It should be noted that the origin of the renormalization 
is different from that for the normal metal island
\cite{Schoeller1,Konig,Schoeller2,Golubev} and the 
double-island\cite{Pohjola} where particle-hole GF is symmetric. 

The expression (\ref{eq:current}) is formally equivalent to that obtained within RTA\cite{note5}. 
The validity of RTA, which is developed for the normal metal island\cite{Schoeller1,Konig,Schoeller2}, 
is not obvious when the superconducting correlation exists. 
However, we expect that RTA can be used as a start point of approximation when QP tunneling is the main transport mechanism. 
The expectation is partly justified by the following two results. 
First, Eq. (\ref{eq:current}) reproduces the result of the orthodox theory\cite{Schon} in the limit of small dimensionless junction conductance $\alpha_0$. 
Secondly, in CB regime ($|eV| \ll |\tilde{\Delta_0}|$), Eq. (\ref{eq:current}) 
is approximately expressed as
\begin{equation}
G_{\rm K} (2 \pi)^2 \alpha_{\rm L}^0 \alpha_{\rm R}^0
(2 e \Delta V)^{1/2} V/3 \rho_{\rm I}^0 \tilde{\Delta}_0^2, 
\nonumber
\end{equation}
which is equal to the previous expression obtained by 
Averin and Nazarov\cite{Averin} in CB regime for the odd state 
($|eV| \ll -\tilde{\Delta_0})$
except for the renormalization of the peak position 
\begin{equation}
\tilde{\Delta}_0=\Delta_0+2 \alpha_0 \Delta \ln(2 E_C/\Delta)-
\ln(2 \rho_{\rm I}^0 E_C/\pi \alpha_0)/\rho_{\rm I}^0. 
\nonumber
\end{equation}
As for the CB regime for the even state ($|eV| \ll \tilde{\Delta_0}$), 
our result does not reproduces the previous result
that the inelastic co-tunneling current is zero\cite{Averin}. 
Here we should comment on the limit of the application of RTA. 
RTA takes account of high-order inelastic co-tunneling 
processes where at most one particle-hole excitation is created 
inside the island at a moment\cite{Schoeller1}. 
And thus, the simple application of RTA for the even state, where 
a low-energy particle-hole excitation cannot be created inside the island, 
causes us to count some undesirable processes. 
To improve the approximation in CB regime for the even 
state is outside the scope of this paper.

\section{results and discussions}
\label{sec:results}


In our numerical calculation, we choose the superconducting gap and DOS as $\Delta/E_C=10^{-1}$ and $\rho_{\rm I}^0 E_C=10^3$, respectively. 
The variable $\rho_{\rm I}^0 \Delta=10^2$ is the same order as that of 
the Al island whose volume is $10^4 {\rm nm}^3$. 
We confirm that our numerical calculation reproduces the spectral sum rule 
of the full $c$-field GF 
$
\int \rd \varepsilon {\rm Im}G^R_c(\varepsilon)=-\pi
$
within $0.2\%$ accuracy. 

\begin{figure}[ht]
\epsfxsize=\linewidth
\epsffile{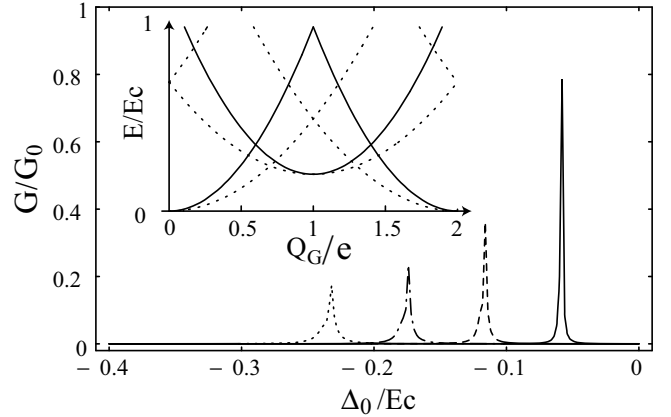}
\caption{
The excitation energy dependence of differential conductance for $\alpha_0=0.1$ (solid line), 0.2 (dashed line) 0.3, (dot-dashed line) and 0.4 (dotted line) at $eV/E_{C}=10^{-4}$. 
Inset: 
A schematic diagram of the total energy with (dashed lines) and without charge fluctuation (solid lines).  
}
\label{fig:conductance}
\end{figure}

Let us first consider the effect of the charge fluctuation on the conductance $G$. 
Using Eq. (\ref{eq:current}), we calculate $G$ as 
\begin{equation}
G/G_{0}
\sim
\alpha_0 (1/2 \rho_{\rm I}^0)^2
/
\{
\tilde{\Delta}_0^2
+
(\pi \alpha_0 /2 \rho_{\rm I}^0)^2
\}, 
\nonumber 
\end{equation}
where 
$G_0=G_{\rm K} (2 \pi)^2 \alpha_0^{\rm L} \alpha_0^{\rm R}/\alpha_0$. 
Figure \ref{fig:conductance} shows the excitation energy dependence of
differential conductance at small bias voltage ($eV/E_C=10^{-4}$) for
various $\alpha_0$.  The width of
the conductance peak is approximately given by $\pi \alpha_0/2 \rho_{\rm I}^0$
inversely proportional to the life-time of the unpaired
electron.  The conductance peak shifts leftward as $\alpha_0$
increases.  The peak position, namely the degeneracy point, is at
$\Delta_0 \sim -2 \alpha_0 \Delta \ln(2 E_C/\Delta)$.  At the
degeneracy point, the gate charge $Q_{\rm G}/e$ is approximately $\Delta/(2 z
E_C)+1/2$, where $z=1/(1+2 \alpha_0 \ln(2 E_C/\Delta))$ is the
renormalization factor.  The charge fluctuation reduces the charging
energy, while it does not affect on the magnitude of the superconducting
gap.  This result is qualitatively different from that of the normal
island, where both of the charging energy and the conductance are
renormalized\cite{Schoeller1}. 

The origin of the shift of the degeneracy point can be understood by
looking at the schematic diagram of gate charge dependence of the
total energy shown in the inset of Fig. \ref{fig:conductance}. The
dashed and the solid line indicate the total energy with and without
charge fluctuation, respectively. 
Here we assume that the renormalization factor is independent of $\Delta_0$. 
One can see that the curvature of the energy curve with the charge
fluctuation is smaller than that without the charge fluctuation since
the charging energy is reduced due to the charge fluctuation.
Thus the degeneracy points are shifted to reduce CB regime for the odd state.

It should be stressed that the renormalization factor can be determined experimentally. 
From the Coulomb oscillation, one can obtain the ratio of
the superconducting gap to the renormalized charging energy
$\Delta/(z E_C)=
(Q^{\rm e}_{\rm G}-Q^{\rm o}_{\rm G})/(Q^{\rm e}_{\rm G}+Q^{\rm o}_{\rm G})$,
where $Q^{\rm o(e)}_{\rm G}$ is the interval corresponding to odd (even) state.
Here the \lq \lq bare" charging energy $E_C$ and the superconducting gap
$\Delta$ can also be obtained experimentally.
For example, one can obtain $E_C$
from the temperature dependence of the conductance peak
for the normal state\cite{Joyez}.
The superconducting gap $\Delta$ can be obtained from the
$I$-$V$ characteristic as discussed below.
Therefore, one can estimate the renormalization factor quantitatively by
analyzing the experimental results. 


\begin{figure}[ht]
\epsfxsize=\linewidth
\epsffile{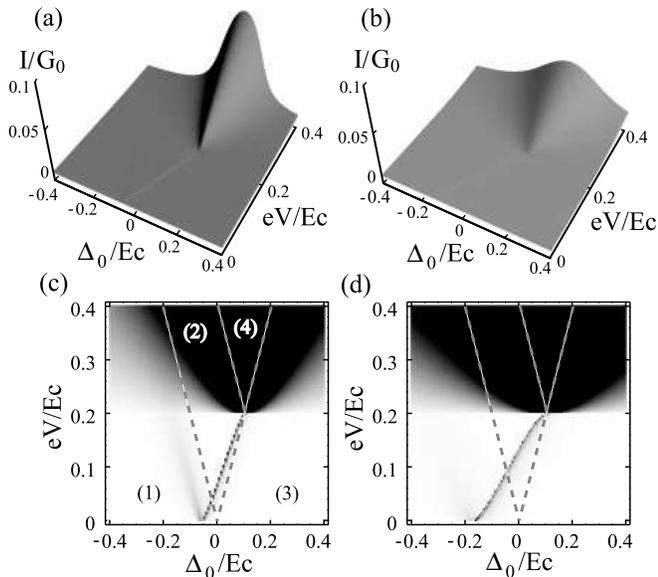}
\caption{
The excitation energy and the bias voltage dependence of 
the tunneling current for (a) $\alpha_0=0.1$ and (b) 0.3. 
The corresponding density plots of differential conductance are shown in panels (c) and (d). 
}
\label{fig:current}
\end{figure}

Figures \ref{fig:current}(a) and
\ref{fig:current}(b) show the excitation and bias voltage dependence
of the tunneling current for $\alpha_0=0.1$ and 0.3, respectively.
The corresponding density plots of differential conductance are shown
in Figs. \ref{fig:current}(c) and \ref{fig:current}(d).  Dotted lines
indicate boundaries of four regions predicted by the orthodox
theory\cite{Schon}: (1) CB regime for the odd state, (2) the unpaired
electron tunneling regime, (3) CB regime for the even state, and (4)
the QP tunneling regime.  The tunneling current shows the plateau in
the region (2). 
As $\alpha_0$ becomes large 
(Figs. \ref{fig:current}(a) and \ref{fig:current}(b)), 
one can see that the structure of the 
tunneling current is smeared by the life-time broadening, 
especially around the QP tunneling regime (4). 
However, the threshold bias voltage where a QP tunneling channel opens 
$eV=2 \Delta$ does not depend on $\alpha_0$ and thus 
one can obtain $\Delta$ experimentally.

The striking feature in Figs. \ref{fig:current}(c) and \ref{fig:current}(d) 
is that at the boundary of CB regime for the even 
state, the conductance peak survives until $eV$ reaches $2 \Delta$. 
This is because the dissipative charge fluctuation 
is blocked due to the superconducting
gap and only the unpaired electron contributes to the
life-time broadening: $\gamma \sim \pi \alpha_0 /2 \rho_{\rm I}^0$.
On the other hand, at the boundary of CB regime for the odd state, 
the conductance peak is gradually smeared. 
This is because the dissipative charge
fluctuation increases with increasing bias voltage, and the life-time
broadening $\gamma$ is proportional to the number of QP states:
$\gamma \sim \pi \alpha_0 \sqrt{\Delta \ eV}$. 

The renormalization effect is expected to be weakened by applied bias voltage 
because $eV$ gives the low energy cut-off for the 
renormalization factor\cite{Schoeller2}.  
The reduction of the renormalization effect can be deduced 
by measuring the peak position of the conductance 
at the boundary of CB regime for the even state 
(Figs. \ref{fig:current}(c) or \ref{fig:current}(d)). 
In the limit of zero bias voltage, the position of the conductance peak 
deviates from that predict by the orthodox theory(dotted line). 
As the bias voltage increases, the position of the peak 
approaches the dotted line 
and they meet at $eV =2 \Delta$. It means that 
the renormalization effect is weakened with increasing bias voltage. 

\section{summary}
\label{sec:summary}

In conclusion, we have theoretically investigated the transport
properties of NSN transistor showing the parity effect and the quantum
fluctuation of charge.  We considered the quantum fluctuation between the 
even and the odd state caused by the inelastic resonant
tunneling precess.  
We found that the charge fluctuation causes the
renormalization of the charging energy, and that CB regime for the odd
state is reduced.  
The renormalization factor can be obtained experimentally. 
The renormalization effect is weakened by applied bias voltage. 
At the boundary of CB regime for the even state, the conductance peak is robust
against the charge fluctuation because only 
one unpaired electron can contribute to the life-time broadening of the charge state.  
On the contrary, at the boundary of CB regime for the odd state, the
conductance peak is smeared since the dissipative charge fluctuation 
increases with increasing applied bias voltage.

\section*{ACKNOWLEDGMENTS}

We would like to thank Y. Isawa, Y. Takane and A. Kanda for valuable discussions. 
One of us (H. I) is supported by MEXT, Grant-in-Aid for Encouragement of Young Scientists, 13740197, 2001. 

\appendix
\section{functional integral on the closed time-path}
\label{appendix:functionalintegration}

In this Appendix, we demonstrate the method to calculate 
the functional integral defined on the closed time path $C$. 
For generality, we consider the following action
\begin{eqnarray}
S
&=&
\int_C \rd t
\{
c(t)^* (\ri \hbar \, \partial_t - \varepsilon) c(t)
+J(t)^* c(t)+h.c.
\},
\end{eqnarray}
where $c$ is a Grassmann variable. 
A complex variable $J$ is a source field and is zero on $C_{\tau}$. 
In order to utilize the time-slicing 
technique on $C$\cite{Okumura}, 
we introduce the cut-off time $t_0 >0$ which restricts the 
range of $C_{\pm}$ from $-t_0$ to $t_0$
and we take the limit as $t_0 \rightarrow \infty$ after all calculations. 
We divide contours $C_{+}$, $C_{-}$ and $C_{\tau}$ into $N$ pieces, 
respectively. 
The discretized time is $t_n=\sum_{i=1,n} \epsilon_i -t_0$ where
$\epsilon_n=2 t_0/N$ for $n=1,\cdots,N$, 
$\epsilon_n=-2 t_0/N$ for $n=N+1,\cdots,2 N$ 
and 
$\epsilon_n=-\ri \hbar \beta/N$ for $n=2 N+1,\cdots,3 N$. 
The action is written in the discretized form as
$
\sum_{i,j=1}^{3N} c^*_i A_{ij} c_j
+\sum_{i=1}^{3 N}  \epsilon_i J^*_i c_i
+h.c.,
$
where
$A_{ij}=\ri \hbar \, \delta_{ij}-
(\ri \hbar+\epsilon_i \varepsilon) \, \delta'_{i \; j+1}$. 
$\delta'_{i \; j+1}$ is the Kronecker's delta $\delta_{i \; j+1}$ 
for $j=1,\cdots,3N-1$ and
$-\delta_{i \; 1}$  for $j=3N$. 
The integration is performed 
in the same way as the imaginary time path-integral\cite{Negele-Orland}. 
By taking the limit as $t_0  \rightarrow \infty$ after 
taking the continuous limit $N \rightarrow \infty$, 
we obtain the generating functional as
\begin{eqnarray}
Z
&=&
Z_0
\exp
\left(
\frac{1}{\ri \hbar}
\int_C \rd t \rd t' J^*(t) g(t,t') J(t')
\right). 
\label{eqn:z}
\end{eqnarray}
Here 
$
Z_0=\exp {\rm Tr} \ln \left( g^{-1} \right)
$
becomes the partition function for fermion, 
$
1+{\rm e}^{-\beta \varepsilon},
$
because the contribution from $C_{+}$ and that from $C_{-}$ chancel 
each other and only the contribution from $C_{\tau}$ remains. 

Though GF in Eq. (\ref{eqn:z}) is defined on $C$, 
in the practical calculations GF defined on $C_{+}+C_{-}$ 
is needed. 
If $t>t'$ with respect to the contour $C_{+}+C_{-}$, 
$g(t,t')$ is written with the time projected on the real axis as
\begin{equation}
f^{+}(\varepsilon)
\exp
(
\varepsilon(t-t')/(\ri \hbar)
)
/(\ri \hbar),
\label{eqn:gsingle1}
\end{equation}
where $f^{+}(\varepsilon)$ is defined with the Fermi function 
$f^{-}(\varepsilon)=1/({\rm e}^{\beta \varepsilon}+1)$
as
$f^{+}(\varepsilon)=f^{-}(\varepsilon) \, {\rm e}^{\beta \varepsilon}$.
In the opposite case, $t<t'$, $g(t,t')$ is written as
\begin{equation}
-f^{-}(\varepsilon)
\exp
(
\varepsilon(t-t')/(\ri \hbar)
)
/(\ri \hbar).
\label{eqn:gsingle2}
\end{equation}

\section{first order perturbation theory}
\label{appendix:charge1st}

In this Appendix, we demonstrate that the lowest order 
perturbation theory gives the 
divergent average charge at the degeneracy point
within our formulation. 
For simplicity, we consider 
the case of the equilibrium state and in the limit of zero temperature. 
In this case, the imaginary-time formalism\cite{Negele-Orland} is convenient
for calculations and 
the grand canonical potential is evaluated perturbatively 
in the same way as Sec. \ref{sec:model}. 
The first order term of the grand canonical potential
$\Omega^{(1)}$ becomes the similar form as $W^{(1)}$
and consists of the thermal GF for $c$-field, $d$-field and 
the particle-hole given by 
$1/(\ri \omega_l-\Delta_0)$,
$2/(\ri \omega_l)$\cite{Spencer}
and 
$\alpha(\ri \nu_n)$, respectively. 
Here, $\omega_l$ and $\nu_n$ are 
the fermion and the boson Matsubara frequency, respectively. 
\begin{eqnarray}
\Omega^{(1)}
&=&
\frac{1}{\beta^2}
\sum_{l,n}
\frac{1}{\ri \nu_n+\ri \omega_l-\Delta_0}
\,
\frac{2}{\ri \omega_l}
\,
\alpha(\ri \nu_n)
\nonumber \\
&=&
{\rm tanh} \left( \frac{\Delta_0}{2 T} \right)
P
\int \rd \varepsilon
\frac{-\frac{1}{\pi} {\rm Im} \left[ \alpha(\varepsilon+\ri 0) \right] }
{\Delta_0-\varepsilon}
N^{-}(\varepsilon), 
\nonumber
\end{eqnarray}
where $P$ denotes the Cauchy's principal value integral and 
$N^{-}(\varepsilon)=1/({\rm e}^{\beta \varepsilon}-1)$
is the Bose distribution function. 
Here the numerator of the integrand is the spectral function of
the particle-hole GF written as $\alpha_0 \rho(\varepsilon-\Delta)$ where
$\rho$ is given by Eq. (\ref{eqn:spectralfunction}). 
The correction of the average charge is evaluated by
the derivative of $\Omega^{(1)}$ 
in terms of the excitation energy\cite{Schoeller1}.
In the limit of zero temperature, 
it becomes as
$\partial \Omega^{(1)}/\partial \Delta_0 
\sim 
\frac{1}{2} \alpha_0
{\rm sgn}(\Delta_0) \,
\partial R(\Delta_0-\Delta)/\partial \Delta_0$
where $R$ is given by Eq. (\ref{eq:self-energy}). 
As a result, in the CB regime for even state 
$0<\Delta_0 \ll E_C$, 
we can see that the correction diverges as
$\sim \alpha_0 \pi \sqrt{\Delta/(2 \Delta_0)}/2$,
where we omit the contribution form the last term of Eq. 
(\ref{eq:self-energy}), which is negligible
when we consider the life-time effect. 
For normal state, our formulation reproduces the 
well known result, the log-divergence\cite{Schon_text}. 
The divergence for NSN transistor 
is strong as compared with the log-divergence.

\section{green functions}
\label{appendix:greenfunction}

In this Appendix we calculate GFs. 
Before proceeding to the calculation of each GF, 
we demonstrate the method to obtain GF by solving
the differential equation
\begin{eqnarray}
g^{-1}(t,t')
&=&
(\ri \hbar \, \partial_t - \varepsilon) \, \delta(t,t'),
\label{eqn:differntialeq}
\\
g(t,-t_0 \in C_{+})&=&-g(t,-\ri \hbar \beta-t_0), 
\label{eqn:boundaryA}
\end{eqnarray}
as complementary to the method demonstrated 
in Appendix \ref{appendix:functionalintegration}\cite{note2}. 
Here the definition of $t_0$ is given in Appendix. 
\ref{appendix:functionalintegration}. 
First, we calculate GF defined on $C_{\tau}$, viz. $t,t' \in C_{\tau}$. 
In this case, the $\delta$-function formally
satisfies the relation
$
\int^{-\ri \hbar \beta}_0 \rd t \delta(t)=1
$. 
By using the anti-periodic 
boundary condition Eq. (\ref{eqn:boundaryA}), we can solve 
Eq. (\ref{eqn:differntialeq}) in the same way as the thermal GF
\cite{Negele-Orland}. 
From the solution, we can show the following relation as 
\begin{eqnarray}
g(-t_0 \in C_{\mp}, -t_0 \in C_{\pm})
=
\pm
f^{\pm}(\varepsilon)/(\ri \hbar),
\label{eqn:condition1}
\end{eqnarray}
where, we use conditions 
$g(-t_0 \in C_{-}, -t_0 \in C_{+})=-g(-t_0 \in C_{-}, -t_0- \ri \hbar \beta)$
and
$g(-t_0 \in C_{+}, -t_0 \in C_{-})=-g(-t_0- \ri \hbar \beta, -t_0 \in C_{-})$. 
Second, we calculate GFs defined on the real axis\cite{note3}. 
By projecting the time defined on $C_{\pm}$, 
onto the real axis, $g(t,t')$ is projected 
as $g^{\pm \pm}(t,t')$ for $t,t' \in C_{\pm}$
and 
as $g^{\pm \mp}(t,t')$ for $t \in C_{\pm}$ and $t' \in C_{\mp}$. 
It should be careful that the arguments $t$ and $t'$ of GFs with superscripts
are the time projected onto the real axis. 
These four GFs satisfy the following differential equations as 
\begin{eqnarray}
& &
(\ri \hbar \partial_t - \varepsilon)
\left \{
\begin{array}{ccc}
g^{\pm \pm}(t,t') &=& \pm \delta(t-t')  \\
g^{\pm \mp}(t,t') &=& 0,
\end{array}
\right.
\nonumber
\end{eqnarray}
where the $\delta$-function is defined on the real axis. 
These differential equations can be solved by using 
Eq. (\ref{eqn:condition1}) and matching conditions,
$g^{s \pm}(t, t_0)=g^{s \mp}(t, t_0)$
and
$g^{\pm s}(t_0, t')=g^{\mp s}(t_0,t')$
where $s=+$ or $-$. 
The resulting GFs are
\begin{eqnarray}
& &
\left \{
\begin{array}{ccc}
g^{++}(t,t') &=& 
g^{-+}(t,t') \, \theta(t-t')+
g^{+-}(t,t') \, \theta(t'-t)
\\
g^{--}(t,t') &=& 
g^{-+}(t,t') \, \theta(t'-t)+
g^{+-}(t,t') \, \theta(t-t')
\\
g^{\mp \pm}(t,t') &=&
\pm f^{\pm}(\varepsilon)
\exp
\left(
\varepsilon (t-t')/(\ri \hbar)
\right)/(\ri \hbar).
\\ 
\end{array}
\right.
\nonumber
\end{eqnarray}
The results are equivalent to Eqs. (\ref{eqn:gsingle1}) and 
(\ref{eqn:gsingle2}). 

The four GFs are components of $2 \times 2$ GF in the 
{\it single time representation}\cite{Chou}. 
It is known that this representation includes the redundancy 
which can be removed by the Keldysh rotation\cite{Keldysh,Chou}. 
After the Keldysh rotation, we obtain $2 \times 2$ 
GF in the physical representation (see Eq. (\ref{eqn:physical})). 
In the following discussions, we summerize 
the retarded and the Keldysh components of $2 \times 2$ GF 
in the physical representation. 
The advanced component is obtained by taking 
the complex conjugate of the retarded component in the energy space.

The differential equation Eq. (\ref{eqn:gk}) can be solved in the same way as 
above example: 
\begin{eqnarray}
g^R_{{\rm r} k}(\varepsilon) &=& 
1/(\varepsilon+\ri \eta-\varepsilon_{{\rm r} k}),
\label{eq:ikR}
\\
g^K_{{\rm r} k}(\varepsilon) &=& -2 \ri \pi
\tanh \left( \frac{\varepsilon}{2 T} \right)
\delta (\varepsilon-\varepsilon_{{\rm r} k}),
\label{eq:ikK}
\end{eqnarray}
where, $\eta$ is a positive small value and 
the $\delta$-function in the energy space is defined as
$
\delta(\varepsilon)=\frac{1}{\pi}
\frac{\eta}{\varepsilon^2+\eta^2}.
$

By using Eqs. (\ref{eq:ikR}) and (\ref{eq:ikK}), 
we can calculate the retarded and the Keldysh 
components of particle-hole GF defined by Eq. (\ref{eqn:gph}). 
In the following, we fix the relative coordinate of the phase difference
as $\varphi_{\Delta}(t)=0$, and the center-of-mass coordinate 
as $\varphi_c(t)=eVt/\hbar$. 
The loop diagram can be calculate in the standard way\cite{note4}:
\begin{eqnarray}
&\alpha_{\rm r}^R&(\varepsilon)
=
N_{\rm ch} T_{\rm r}^2
\sum_{k,k'}
\int \frac{\rd \varepsilon'}{2 \ri \pi}
\nonumber \\ &\times&
\frac{
g_{{\rm r} k}^R(\delta \varepsilon^{\rm r}+\varepsilon') 
g_{{\rm I} k'}^K(\varepsilon')
+
g_{{\rm r} k}^K(\delta \varepsilon^{\rm r}+\varepsilon') 
g_{{\rm I} k'}^A(\varepsilon')
}{2}
\label{eqn:loopR}
\\
&=&
-\ri \pi
\alpha^0_{\rm r}
\rho(\delta \varepsilon^{\rm r}-\Delta)
\label{eqn:alphaR}, 
\\
&\alpha_{\rm r}^K&(\varepsilon)
=
N_{\rm ch} T_{\rm r}^2
\sum_{k,k'}
\int \frac{\rd \varepsilon'}{2 \ri \pi}
\nonumber \\ &\times&
\frac{
g_{{\rm r} k}^K(\delta \varepsilon^{\rm r}+\varepsilon') 
g_{{\rm I} k'}^K(\varepsilon')
-
g_{{\rm r} k}^C(\delta \varepsilon^{\rm r}+\varepsilon') 
g_{{\rm I} k'}^C(\varepsilon')
}{2}
\label{eqn:loopK}
\\
&=&
- 2 \ri \pi
\alpha^0_{\rm r}
\rho(\delta \varepsilon^{\rm r}-\Delta)
\coth
\left(
\frac{\delta \varepsilon^{\rm r}}{2T}
\right)
\label{eqn:alphaK}, 
\end{eqnarray}
where the spectral function $\rho(\varepsilon)$ is defined by Eq. 
(\ref{eqn:spectralfunction}). 

By solving the differential equations (\ref{eqn:GFc}) and (\ref{eqn:GFd}), 
we obtain GFs for $c$- and those for $d$-field: 
\begin{eqnarray}
g^R_{\phi}(\varepsilon) &=& 2/(\varepsilon+\ri \eta),
\label{eq:dR}
\\
g^K_{\phi}(\varepsilon) &=& 0,
\label{eq:dK}
\\
g^R_c(\varepsilon) &=& 1/(\varepsilon+\ri \eta-\Delta_0),
\label{eq:rc}
\\
g^K_c(\varepsilon) &=& -2 \ri \pi
\tanh \left( \frac{\varepsilon}{2 T} \right)
\delta (\varepsilon-\Delta_0).
\label{eq:Kc}
\end{eqnarray}

The retarded and the Keldysh component of self-energy for $c$-field 
Eq. (\ref{eqn:selfc}) are calculated by using Eqs. 
(\ref{eqn:alphaR}), (\ref{eqn:alphaK}), 
(\ref{eq:dR}), and (\ref{eq:dK}): 
\begin{eqnarray}
& &\Sigma_{\rm r}^R(\varepsilon)
=
\int
\frac{\rd \varepsilon'}{2 \pi}
\frac{\ri \, \alpha_{\rm r}^K(\varepsilon')}
{\varepsilon+\ri \eta -\varepsilon'},
\label{eqn:selfenergyR}
\\
& &
\Sigma_{\rm r}^K(\varepsilon)
=
\alpha_{\rm r}^R(\varepsilon)-\alpha_{\rm r}^A(\varepsilon)
=
2 \alpha_{\rm r}^R(\varepsilon),
\label{eqn:selfenergyK}
\end{eqnarray}
where we utilize the similar expression as
Eqs. (\ref{eqn:loopR}) and (\ref{eqn:loopK}).

The expressions for full $c$-field GF is obtained by solving the 
inverse Dyson equation Eq. (\ref{eqn:fullgc}). 
By projecting the time on $C_{\pm}$ onto the real axis, 
and performing the Fourier transformation, 
we obtain the matrix Dyson equation as 
$
\tilde{G}_c(\varepsilon)
=
\tilde{g}_c(\varepsilon)
-
\tilde{g}_c(\varepsilon)
\mtau^1
\tilde{\Sigma}_c(\varepsilon)
\mtau^1
\tilde{G}_c(\varepsilon)
$. 
The matrix Dyson equation is solved easily: 
\begin{eqnarray}
G_c^R(\varepsilon)
&=&
1/(
\varepsilon+\ri \eta-\Delta_0
-\Sigma_c^R(\varepsilon))
\label{eqn:cGC},
\\
G_c^K(\varepsilon)
&=&
G_c^R(\varepsilon)
\left \{
\Sigma_c^K(\varepsilon)
-2 \ri \, \eta \tanh \left( \frac{\varepsilon}{2 T} \right)
\right \}
G_c^A(\varepsilon), 
\label{eqn:cGK}
\end{eqnarray}
where we use the definition of $\delta$-function in the energy space. 
Taking the limit $\eta \rightarrow 0$, we obtain the final form 
for full $c$-field GF.  


%
%
%

\begin{references}
%
\bibitem{Falci}
G. Falci, G. Sch\"on, and G. T. Zimanyi, 
Phys. Rev. Lett. {\bf 74}, 3257 (1995). 
%
\bibitem{Schoeller1}
H. Schoeller, and G. Sch\"on, 
Phys. Rev. B {\bf 50}, 18436 (1994). 
%
\bibitem{Konig}
J, K\"onig, H. Schoeller, G. Sch\"on, and R. Fazio, in 
{\it Quantum Dynamics of Submicron Structures}, 
edited by H. A. Cerdeira {\it et al.} 
(Kluwer, Dordrecht, 1995),
pp. 221-239. 
%
\bibitem{Schoeller2}
H. Schoeller, in 
{\it Mesoscopic Electron Transport}, 
edited by L. L. Sohn {\it et al.} 
(Kluwer, Dordrecht, 1997), 
pp. 291-330.
%
\bibitem{Golubev}
D. S. Golubev, J. K\"onig, H. Schoeller, G. Sch\"on, and A. D. Zaikin, 
Phys. Rev. B {\bf 56}, 15782 (1997). 
%
\bibitem{Joyez}
P. Joyez, V. Bouchiat, D. Esteve, C. Urbina, and M. H. Devoret, 
Phys. Rev. Lett. {\bf 79}, 1349 (1997). 
%
\bibitem{Chouvaev}
D. Chouvaev, L. S. Kuzmin, D. S. Golubev, and A. D. Zaikin, 
Phys. Rev. B {\bf 59}, 10599 (1999). 
%
\bibitem{RBT}
D. C. Ralph, C. T. Black, J. M. Hergenrother, J. G. Lu and M. Tinkham, 
in the same book as Ref. \cite{Schoeller2}, 
pp. 447-467, 
and references therein. 
%
\bibitem{Fazio}
R. Fazio and G. Sch\"on, 
in the same book as Ref. \cite{Schoeller2}, 
pp. 407-446. 
%
\bibitem{Utsumi1}
Y. Utsumi, M. Hayashi and H. Ebisawa, 
J. Phys. Soc. Jpn. {\bf 69}, 2739 (2000).
%
\bibitem{Hekking1}
F. W. J. Hekking, L. I. Glazman, K. A. Matveev and R. I. Shekhter, 
Phys. Rev. Lett. {\bf 70}, 4138 (1993).
%
%
%
\bibitem{Hekking2}
F. W. J. Hekking and Yu. V. Nazarov, 
Phys. Rev. Lett. {\bf 71}, 1625 (1993). 
%
\bibitem{Isawa}
Y. Isawa and H. Horii, 
J. Phys. Soc. Jpn. {\bf 69}, 655 (2000).
%
\bibitem{Spencer}
H. J. Spencer,
Phys. Rev. {\bf 167}, 434 (1968). 
%
\bibitem{Keldysh}
L. V. Keldysh, 
Sov. Phys. JETP {\bf 20}, 1018 (1965). 
%
\bibitem{Chou}
K.-C. Chou, Z.-B. Su, B.-L. Hao and L. Yu, 
Phys. Rep. {\bf 118}, 1 (1985).
%
\bibitem{Kamenev}
A. Kamenev and A. Andreev, 
Phys. Rev. B {\bf 60}, 2218 (1999).
%
\bibitem{Gutman1}
D. B. Gutman and Y. Gefen,
Phys. Rev. B {\bf 64}, 205317 (2001). 
%
%
%
\bibitem{Babichenko}
V. S. Babichenko and A. N. Kozlov, 
Solid State Commun. {\bf 59}, 39, (1986). 
%
\bibitem{Kiselev}
M. N. Kiselev and R. Oppermann, 
Phys. Rev. Lett. {\bf 85}, 5631, (2000). 
%
\bibitem{Pohjola}
T. Pohjola, J. K\"onig, H. Schoeller and G. Sch\"on, 
Phys. Rev. B {\bf 59}, 7579 (1999).
%
\bibitem{note5}
Exactly speaking, our result is formally equivalent 
to that of Ref. \cite{Schoeller1}, however 
not equivalent to  the result of Ref. \cite{Isawa}. 
In Ref. \cite{Isawa}, an excess sequential tunneling process 
is proposed, in addition to the resonant tunneling 
process\cite{Schoeller1}. 
%
\bibitem{Schon}
G. Sch\"on, J. Siewert and A. D. Zaikin, 
Physica B {\bf 203}, 340 (1994).
%
\bibitem{Averin}
D. V. Averin, and Yu. V. Nazarov, 
Phys. Rev. Lett. {\bf 69}, 1993 (1992).
%
\bibitem{Okumura}
K. Okumura and Y. Tanimura,
Phys. Rev. E {\bf 53}, 214 (1996). 
%
\bibitem{Negele-Orland}
J. W. Negele, and H. Orland,
{\it Quantum Many-Particle Systems}
(Addison-Wesley Publishing Company, Redwood City, California, 1988).
%
\bibitem{Schon_text}
G. Sch\"on, in
{\it Quantum Transport and Dissipation}, 
edited by T. Dittrich {\it et al.}
(Wiley-VCH, Weinheim, 1998),
pp. 149-212. 
%
\bibitem{note2}
Though this is the basic knowledge for Keldysh formalism, 
we dare to explain the method to solve 
Eq. (\ref{eqn:differntialeq}).
Because there is no reasonable text book 
which explains the method to solve the differential equation
defined not on the Keldysh contour $C_{+}+C_{-}$ 
but on the contour $C$. 
%
\bibitem{note3}
Once the boundary condition at $t=-t_0$ is determined as 
Eq. (\ref{eqn:condition1}), the following process 
becomes equal in the case of the Keldysh 
contour $C_{+}+C_{-}$. 
%
\bibitem{note4}
The method to calculate the loop diagram is presented in the review
[J. Rammer, and H. Smith, Rev. Mod. Phys. {\bf 58}, 323 (1986)]. 
For example, we calculate the retarded and the Keldysh components of 
the loop diagram $\Pi(t,t')=g_1(t,t') \, g_2(t',t)$. 
$\Pi^R(t,t')$ and $\Pi^K(t,t')$ can be 
obtained by calculating the trace of $2 \times 2$ matrices
${\rm Tr}[\mtau^0 \tilde{g}_1(t,t') \mtau^1 \tilde{g}_2(t',t)]/2$
and 
${\rm Tr}[\mtau^0 \tilde{g}_1(t,t') \mtau^0 \tilde{g}_2(t',t)]/2$, 
respectively. 
By using the relation $g_1^{R(A)}(t,t') \, g_2^{R(A)}(t',t)=0$, 
we obtain the final expressions: 
$\Pi^R(t,t')=\{ g_1^R(t,t') g_2^K(t',t)+g_1^K(t,t') g_2^A(t',t)\}/2$
and
$\Pi^K(t,t')=\{ g_1^K(t,t') g_2^K(t',t)-g_1^C(t,t') g_2^C(t',t)\}/2$. 
%
\end{references}
\end{document}